\documentclass[prd,twocolumn,preprintnumbers,superscriptaddress,nofootinbib]{revtex4}
\usepackage[a4paper, hdivide={1.91cm,,1.165cm}, vdivide={1.83cm,,3.6cm}]{geometry}

\usepackage{amstext,amssymb}
\usepackage{amsmath}
\usepackage{graphicx}
\usepackage[hyperfootnotes=true]{hyperref}
\usepackage{xspace}
\usepackage{color}
\usepackage{units}
\usepackage{slashed} 

\def \vec#1{{\boldsymbol{#1}}}

\begin{document}

\title{Dark matter, lepton and baryon number,
and left-right symmetric theories}
\author{Sudhanwa Patra}
\email{sudha.astro@gmail.com}
\affiliation{Center of Excellence in Theoretical and Mathematical Sciences, \\
 \hspace*{-0.0cm} Siksha 'O' Anusandhan University, Bhubaneswar-751030, India}
\begin{abstract}

A lepto-baryonic left-right symmetric theory is considered  along with pointing out stable dark matter candidates 
whose stability is ensured automatically where leptons and baryons are defined as local gauge symmetries. 
These theories are generally anomalous and the possible 
gauge anomaly free solutions for these theories are presented. It is found that the neutral component of fermion triplets can be 
a viable dark matter candidate originally introduced for gauge anomaly cancellation. The other dark matter 
possibilities within this lepto-baryonic left-right symmetric theory are also presented.

\end{abstract}
\maketitle 
\section{Introduction} 
Some of the unsolved puzzles of the Universe like tiny neutrino masses, matter-antimatter asymmetry, 
dark matter and dark energy, and coupling unification of three fundamental interactions have brought the 
realization that, the Standard Model (SM) of particle physics is unlikely to be the final theory and needs to be extended beyond its 
boundaries. Dark matter is the one such fascinating problem of modern cosmology contributing to 26.8\% 
of the total energy budget in our Universe~\cite{Ade:2013zuv}. Although we have indirect gravitational 
evidence of dark matter from the galaxy rotation curve, gravitational lensing, and large-scale structure 
of the Universe, direct detection still remains a mystery. It is highly desirable to identify the 
particle nature of dark matter as well its other properties. The only information about dark matter--at present--
is its relic abundance which is given by $\Omega_{\rm DM}h^2=0.119$~\cite{Ade:2013zuv} whereas SM of 
particle physics cannot accommodate any candidate for dark matter. This motivates the particle physics 
community to propose a beyond SM framework capable of offering numerous possibilities 
of new physics which need to be explored with their experimental verifications. We aim here 
to present a beyond SM framework-- a different class of left-right theories--to list the 
successful dark matter candidates where the stability of dark matter arises naturally.

Among all the possibilities, the left-right symmetric models based on the gauge group 
$SU(2)_L \times SU(2)_R \times U(1)_{B-L} \times SU(3)_C$ \cite{Mohapatra:1974gc, Pati:1974yy, Senjanovic:1975rk,Senjanovic:1978ev,
Mohapatra:1979ia,Mohapatra:1980yp,Pati:1973uk,Pati:1974vw} seems quite promising and is picked often 
for explaining the origin of neutrino masses and parity violation at the electroweak scale.
The underlying mechanisms 
for originating light neutrino masses in these manifest LRSMs are the type-I plus type-II 
\cite{Mohapatra:1979ia,Mohapatra:1980yp,Deshpande:1990ip} 
seesaw mechanisms where the parity and $SU(2)_R$ gauge symmetry are spontaneously broken at the same scale.
Apart from this, when left-right symmetry breaks at a few TeV, it delivers 
rich collider phenomenology including the recent diboson excess reported by ATLAS and some low-energy 
processes like neutrinoless double beta decay which will confirm the lepton number violation in 
nature if experimentally observed. A good number of attempts have been made to address 
the issue of dark matter in these manifest left-right symmetric models as well \cite{Heeck:2015qra,Garcia-Cely:2015quu}
\footnote{See some important studies in TeV scale left-right symmetric models in connection 
to LHC studies, Neutrinoless double beta decay and lepton flavor violation 
\cite{Dev:2015pga,Deppisch:2015cua,Patra:2015bga,Deppisch:2014qpa,Tello:2010am,Das:2012ii,Barry:2013xxa,
Awasthi:2013ff,Chakrabortty:2012mh,Dobrescu:2015qna,Brehmer:2015cia,Dev:2013vxa,Chen:2013fna,Dev:2013oxa}.}.

However, a different class of left-right symmetric model is considered here in which 
the leptons and baryons are defined as local gauge symmetries. This class of 
left-right symmetric model is termed "the lepto-baryonic left-right symmetric model (LB-LRSM)" and is based on 
the gauge group $SU(2)_L \times SU(2)_R \times U(1)_{B} \times U(1)_L \times SU(3)_C$ \cite{He:1989mi,Duerr:2013opa,Patra:2015vmp}
\footnote{The gauge theory for baryons and lepton has been proposed in Ref.\cite{Duerr:2013dza}}. 
The rank of this LB-LRSM being six, it cannot be embedded in a known grand unified theory (GUT) model like $SU(5)$ or 
$SO(10)$ GUT. Any theory which can be embedded in GUT is constrained by the proton decay constraints 
to have the cut-off scale $> 10^{15.6}$ GeV . Interestingly this limit is not applicable to 
LB-LRSM and thus, the unification scale can be lower significantly which might lead to striking signatures. 

In minimal left-right symmetric models the relic density and indirect detection 
including Sommerfeld enhancement push the dark matter mass beyond a few TeV making it unsuitable for a LHC probe. But good enough from 
a collider prospective, in LB-LRSM, the singlet fermion dark matter lies at a low scale, which is studied in a recent work 
\cite{Patra:2015vmp}.
However, this paper focuses on LB-LRSM and scrutinizes all dark matter candidates which can be accommodated in it successfully. 
The plan of the paper is as follows. In Sec. II, the LB-LRSM with one-stage and two-stage breaking to the SM 
gauge theory is studied. In Sec. III, all viable dark matter candidates that can arise naturally in LB-LRSM are discussed. The 
result is summarized towards the end, and the interactions for LB-LRSM are provided in the Appendix.
\section{Lepto-Baryonic Left-Right Models}
The lepto-baryonic left-right symmetric models (LB-LRSM) are based on the basic gauge group 
\cite{He:1989mi,Duerr:2013opa,Patra:2015vmp}
\begin{align}
\mathcal{G}^{LB}_{LR}\equiv SU(2)_L \times SU(2)_R \times U(1)_{B} \times U(1)_{L}\, .
\label{eq:LB-LRSM}
\end{align}
Here we omit the $SU(3)_C$ structure for simplicity. The leptons, and baryons are defined as 
separate local gauge symmetries of the theory. The electric charge is defined as 
\begin{align}
Q=T_{3L}+T_{3R}+\frac{B-L}{2}\, .
\end{align}
Under the LB-LRSM gauge group, the usual quarks and leptons transform as
\begin{eqnarray}
&&q_{L}=\begin{pmatrix}u_{L}\\
d_{L}\end{pmatrix}\equiv[2,1,1/3,0]\,, ~ q_{R}=\begin{pmatrix}u_{R}\\
d_{R}\end{pmatrix}\equiv[1,2,1/3,0]\,,\nonumber \\
&&\ell_{L}=\begin{pmatrix}\nu_{L}\\
e_{L}\end{pmatrix}\equiv[2,1,0,1] \, , ~ \quad \ell_{R}=\begin{pmatrix}\nu_{R}\\
e_{R}\end{pmatrix}\equiv[1,2,0,1] \, . \nonumber
\end{eqnarray}
It is found that these leptonic and baryonic currents are quantum mechanically anomalous. To have a consistent 
anomaly-free left-right symmetric theory one needs to introduce additional fermion degrees of freedom having nonzero 
lepton and baryon charges. There are many ways one can construct an anomaly-free theory of LB-LRSM which has been pointed out 
in the Appendix. A simple extension by additional fermion triplets $\Sigma_L\oplus \Sigma_R$ is considered to serve the purpose, 
\begin{eqnarray}
&&\Sigma_{L}=\begin{pmatrix}
  \Sigma^0_{L,R}  & \sqrt{2} \Sigma^+_{L,R}  \\
  \sqrt{2} \Sigma^-_{L,R} & -\Sigma^0_{L,R}
 \end{pmatrix} \equiv [3,1,-3/4,-3/4] \, , \nonumber \\
&&\Sigma_{R}=\begin{pmatrix}
  \Sigma^0_{R}  & \sqrt{2} \Sigma^+_{R}  \\
  \sqrt{2} \Sigma^-_{R} & -\Sigma^0_{R}
 \end{pmatrix} \equiv [1,3,-3/4,-3/4] \, .
\end{eqnarray}
The motivation behind introducing fermion triplets is twofold:
\begin{itemize}
\item It cancels a nontrivial gauge anomaly making the lepto-baryonic Left-Right symmetric models 
      anomaly-free theories. 
\item The neutral component of these fermion triplets can be stable cold dark matter where 
      the stability is ensured automatically.
\end{itemize}
The spontaneous symmetry breaking of LB-LRSM can be done in different ways leading to interesting 
phenomenology. However, in this work we consider only two-stage and one-stage breaking of LB-LRSM 
to the SM gauge group discussed in the following subsections.
\subsection{Two-stage symmetry breaking}
The spontaneous symmetry breaking of the lepto-baryonic left-right symmetric models (LB-LRSM), in which the symmetry breaking 
of LB-LRSM to the SM gauge group occurs via two-step breaking, is presented below: 
\begin{align}
\mathcal{G}^{LB}_{LR} \mathop{\longrightarrow}^{\langle S_{LB}\rangle} \mathcal{G}_{LR} 
\mathop{\longrightarrow}^{\langle \Delta_R \rangle} \mathcal{G}_{SM} \mathop{\longrightarrow}^{\langle \phi \rangle} U(1)_{\rm em}
\end{align}
where
\begin{align}
&\mathcal{G}^{LB}_{LR} \equiv SU(2)_L \times SU(2)_R \times U(1)_B\times U(1)_{L} \, , \nonumber \\
&\mathcal{G}_{LR} \equiv SU(2)_L \times SU(2)_R \times U(1)_{B-L}\, , \nonumber \\
&\mathcal{G}_{SM} \equiv SU(2)_L \times U(1)_{Y}\, , \nonumber
\end{align}
omitting the $SU(3)_C$ structure for simplicity. Alternatively, the symmetry-breaking pattern 
for two-stage breaking of LB-LRSM can also be understood as follows
\begin{widetext}
\begin{eqnarray*}
&\hspace*{-2cm} \begin{array}[t]{c} SU(2)_L\\ \{T_L, T_{3L} \} \\ g_L(M_{LB})\end{array} \times 
\begin{array}[t]{c} SU(2)_R\\ \{T_R, T_{3R} \} \\ g_R(M_{LB})\end{array} \times 
\underbrace{\begin{array}[t]{c} U(1)_{B}\\ {\rm B}    \\ g_{B}\end{array} \times 
\begin{array}[t]{c} U(1)_{L}\\ {\rm L}    \\ g_{\ell}\end{array}} \bigotimes
\begin{array}[t]{c} \hspace*{-0.0cm} \mathcal{P} \\  \\ \end{array} &\\
&\hspace*{4.4cm} \downarrow  \langle S_{LB}(1,1,3/2,3/2) \rangle 
          \hspace*{1cm} \{ {\color{red} M_S, M_{Z_B}, M_{Z_{\ell}}} \}& \\
&\hspace*{-2cm}  \begin{array}[t]{c} SU(2)_L\\ \{T_L, T_{3L} \} \\ g_L(M_R)\end{array} \times 
\underbrace{\begin{array}[t]{c} SU(2)_R\\ \{T_R, T_{3R} \} \\ g_R(M_R)\end{array} \times 
\begin{array}[t]{c} U(1)_{B-L}\\ {\rm B-L}    \\ g_{BL}\end{array}} 
\begin{array}[t]{c} 
\\Q=T_{3L}+T_{3R}+\frac{B-L}{2}   \\ 
\end{array} &\\
&\hspace*{4.0cm} \downarrow  \langle \Delta_R(1,3,0,-2)\rangle \hspace*{1cm} 
           \hspace*{1.0cm} \{ {\color{red} M_{W_R},M_{Z_R},M_{\Delta}} \} & \\
&\hspace*{-6.0cm} \underbrace{\begin{array}[t]{c} SU(2)_L\\ \{T_L, T_{3L} \} \\ g\equiv g_L\end{array} \times 
\begin{array}[t]{c} U(1)_Y\\ Y       \\ g^\prime \end{array}} & \\ 
&\hspace*{4.2cm} \downarrow  \langle \phi(1,\frac{1}{2})\rangle \quad 
            \hspace*{3.0cm} \{\mbox{SM spectrum with}\, {\color{red} M_{W},M_{Z}}\}& \\
&\hspace*{2.0cm}\begin{array}[t]{c} U(1)_{\rm em}\\ e \end{array} 
\begin{array}[t]{c}  \\  \hspace*{5.0cm} Q=T_{3L} +Y \end{array} &
\end{eqnarray*}
\end{widetext}
Additional discrete left-right symmetry is imposed ensuring equality between $SU(2)_L$ and $SU(2)_R$ gauge
couplings, i.e, $g_L=g_R$ by either parity $\mathcal{P}$ or charge-conjugation $\mathcal{C}$ symmetry. 

The phenomenologically interesting two-stage breaking of the LB-LRSM can be understood in following ways:
\begin{itemize}
\item At first, the spontaneous symmetry breaking of $U(1)_B \times U(1)_L$ to 
$U(1)_{B-L}$ happens via a scalar $S_{LB}(1,1,3/2,3/2)$ 
singlet under $SU(2)_{L,R}$. The inclusion of the scalar not only does the job of spontaneous symmetry breaking but also 
gives masses to extra fermion triplets though 
$$\lambda_\Sigma \mbox{Tr}\left(\Sigma_L \Sigma_L + \Sigma_R \Sigma_R \right) S_{LB} $$
\item The second stage of symmetry breaking is done as in the usual LR symmetric model where 
the breaking of $SU(2)_R\times U(1)_{B-L}$ to $U(1)_Y$ is achieved by scalar triplets $\Delta_L(3,1,-2)$+$\Delta_R(1,3,-2)$. 
However, this stage of symmetry breaking can be implemented via scalar doublets $H_L(2,1,-1)$+$H_R(1,2,-1)$. 
\end{itemize}

The advantage of two-stage symmetry breaking is that one can make an analogy with the minimal left-right 
symmetric model and its low scale phenomenology. The LB-LRSM with scalar triplets, and scalar bidoublet, along 
with scalar singlet are 
\begin{eqnarray}
&&
\Phi=
\begin{pmatrix} 
\phi_{1}^0     &  \phi_{2}^+ \\
\phi_{1}^-     &  \phi_{2}^0
\end{pmatrix} \equiv[2,2,0,0]\,,\nonumber \\
&&
\Delta_{L} \equiv 
\begin{pmatrix} \delta_{L}^+/\sqrt{2} & \delta_{L}^{++} \\ \delta_{L}^0 & -\delta_{L}^+/\sqrt{2} \end{pmatrix}\equiv[3,1,0,-2]  \,,\nonumber \\
&&
\Delta_{R} \equiv 
\begin{pmatrix} \delta_{R}^+/\sqrt{2} & \delta_{R}^{++} \\ \delta_{R}^0 & -\delta_{R}^+/\sqrt{2} \end{pmatrix}\equiv[1,3,0,-2] \,,\nonumber \\
&&
S_{LB} \equiv [1,1,3/2,3/2]
\end{eqnarray}
With scalar triplets $\Delta_L \oplus \Delta_R$, the light neutrino masses are governed by 
type-I plus type-II seesaw mechanisms. The neural lepton mass matrix is given by
\begin{equation}
	M_\nu= 
	\left(\begin{array}{cc}
		M_L  & M_D   \\
   	M^T_D & M_R
\end{array} \right) \, ,
\label{eqn:numatrix}       
\end{equation}
where the Majorana mass matrix for left-handed (right-handed) neutrinos is 
$M_L= f_L \langle \Delta_L \rangle = f v_L $ ($M_R= f_R \langle \Delta_R \rangle = f v_R$ 
while the Dirac neutrino mass matrix connecting light-heavy neutrinos is $m_D= Y_3 v_1 + Y_4 v_2$. 
This results in a mass formula for light as well as heavy neutrinos as
\begin{equation}
	m_\nu = M_L - m_D  M^{-1}_R m^T_D
	= m_\nu^{II} + m_\nu^I\,,
\label{neutrino-mass}
\end{equation}
The consequences of the LRSM seesaw mechanism to neutrinoless double beta decay, LFV and collider studies 
have been explored recently. 
This will be applicable to LB-LRSM if the symmetry breaking happens at a few TeV 
scale with an additional signature which might discriminate between minimal LRSM 
and LB-LRSM in the future. 

\subsection{One-stage symmetry breaking} 
The symmetry-breaking chain of LB-LRSM, in which LB-LRSM to SM symmetry breaking occurs in 
single stage, is given by 
\begin{align}
\mathcal{G}^{LB}_{LR} \mathop{\longrightarrow}^{\langle H_R\rangle,\langle S_{LB}\rangle} \mathcal{G}_{SM} 
\mathop{\longrightarrow}^{\langle \Phi \rangle, \langle H_{L}\rangle}  U(1)_{\rm em}
\end{align}
where,
\begin{align}
&\mathcal{G}^{LB}_{LR} \equiv SU(2)_L \times SU(2)_R \times U(1)_B \times U(1)_{L} \, , \nonumber \\
&\mathcal{G}_{SM} \equiv SU(2)_L \times U(1)_{Y}\, , \nonumber
\end{align}
omitting the $SU(3)_C$ structure for simplicity. This can also be explained in greater detail as follows:
\begin{widetext}
\begin{eqnarray*}
&\hspace*{-2cm} \begin{array}[t]{c} SU(2)_L\\ \{T_L, T_{3L} \} \\ g_L(M_{LB})\end{array} \times 
\underbrace{\begin{array}[t]{c} SU(2)_R\\ \{T_R, T_{3R} \} \\ g_R(M_{LB})\end{array} \times 
\begin{array}[t]{c} U(1)_{B}\\ {\rm B}    \\ g_{B}\end{array} \times 
\begin{array}[t]{c} U(1)_{L}\\ {\rm L}    \\ g_{\ell}\end{array}} 
\begin{array}[t]{c} \hspace*{-0.0cm}  \\  \\ \end{array} &\\
&\hspace*{2.4cm} \downarrow  \langle S_{LB}\rangle, \langle  H_{R}\rangle 
          \hspace*{1cm} \{ {\color{red} M_{Z_B}, M_{Z_\ell}, M_S, M_H, \cdots} \}& \\
&\hspace*{-6.0cm} \underbrace{\begin{array}[t]{c} SU(2)_L\\ \{T_L, T_{3L} \} \\ g\equiv g_L\end{array} \times 
\begin{array}[t]{c} U(1)_Y\\ Y       \\ g^\prime \end{array}} & \\ 
&\hspace*{4.2cm} \downarrow  \langle \phi(1,\frac{1}{2})\rangle \quad 
            \hspace*{3.0cm} \{\mbox{SM spectrum with}\,~ {\color{red} M_{W},M_{Z}}\}& \\
&\hspace*{2.0cm}\begin{array}[t]{c} U(1)_{\rm em}\\ e \end{array} 
\begin{array}[t]{c}  \\  \hspace*{5.0cm} Q=T_{3L} +Y \end{array} &
\end{eqnarray*}
\end{widetext}
Implementation of spontaneous symmetry breaking can be done via Higgs doublets with $B-L\neq 0$ plus lepto-baryonic $SU(2)$-singlet scalar
without the need for a scalar bidoublet with zero $B$ and $L$ charges. Here the fermion masses and mixing can be explained by introducing 
vectorlike fermions via a universal seesaw mechanism. With the introduction of scalar bidoublet, the symmetry breaking 
can be achieved either with scalar doublets 
or scalar triplets plus a lepto-baryonic $SU(2)$-singlet scalar. The role of the lepto-baryonic $SU(2)-$singlet scalar 
is to give the Majorana mass for lepto-baryons 
needed for gauge anomaly cancellation. 
In the present work, the symmetry breaking of $\mathcal{G}^{LB}_{LR} \to \mathcal{G}_{SM} $  as well as for $\mathcal{G}_{SM} \to U(1)_{em}$ 
 is done with $H_R, H_L$, $S_{LB}$, and $\phi$ contained in a bidoublet scalar. 

We present below one such gauge anomaly-free LB-LRSM comprising of fermions as follows 
\begin{eqnarray}
&&q_{L} \equiv[2,1,1/3,0],\quad \quad  q_{R} \equiv[1,2,1/3,0]\,,\nonumber \\
&&\ell_{L} \equiv[2,1,0,1],\quad \quad \quad \ell_{R} \equiv[1,2,0,1] \,,\nonumber \\
&&\Sigma_{L} \equiv[3,1,-3/4,-3/4],\quad \Sigma_{R} \equiv[1,3,-3/4,-3/4] \, .\nonumber
\end{eqnarray}
while the scalar sector is comprising
\begin{eqnarray}
&&
\Phi=
\begin{pmatrix} 
\phi_{1}^0     &  \phi_{2}^+ \\
\phi_{1}^-     &  \phi_{2}^0
\end{pmatrix} \equiv[2,2,0,0]\,,\nonumber \\
&&
H_L=
\begin{pmatrix} 
h_L^+\\
h_L^0
\end{pmatrix}\equiv[2,1,3/4,-1/4]  \,,\nonumber \\
&&
H_R=
\begin{pmatrix} 
h_R^+  \\
h_R^0
\end{pmatrix}\equiv[1,2,3/4,-1/4] \,,\nonumber \\
&&
S_{LB} \equiv [1,1,3/2,3/2]
\end{eqnarray}
The spontaneous symmetry breaking of LB-LRSM $\mathcal{G}^{LB}_{LR}$ to SM is achieved by 
assigning non-zero vevs to $H_R(1,2,3/4,-1/4)$ and $S_{LB}(1,1,3/2,3/2)$. 
Later, the SM gauge group breaks down to $U(1)_{\rm em}$ by the SM Higgs doublet contained in the LR bidoublet 
as $\Phi(2,2,0,0)$. The left-handed counterpart $H_L(2,1,3/4,-1/4)$ is needed from the left-right symmetry. 
The phenomenology of this kind of LB-LRSM is recently studied in Ref.\cite{Duerr:2013opa} in the context 
of light neutrino masses via the type-III seesaw mechanism. The details of the interaction Lagrangian is presented 
in Appendix.

However, one can assign different $B$ and $L$ charges for $H_L\oplus H_R$ such that the electric charge for  
individual components $h^{+}_{L,R}$ and $h^{0}_{L,R}$ should be $+1$ and $0$, respectively, satisfying 
the electric charge relation
\begin{align}
Q=T_{3L} + T_{3R} + \frac{B-L}{2}  
\end{align}
where, $T_{3L}$ ($T_{3R}$) is the third component of the isospin generator for gauge groups 
$SU(2)_L$ ($SU(2)_R$). The reason for choosing $B=3/4$ and $L=-1/4$ for $H_{L,R}$ is to allow 
the $\ell_{L,R} \Sigma_{L,R} H_{L,R}$ terms for implementation of the type-III seesaw. 

The sub-eV scale of light neutrinos can also be explained via the Dirac neutrino mass $M^D_\nu\equiv \frac{1}{\sqrt{2}} 
\left(Y_3 v_1 + Y_4 v_2\right)$ by suitably adjusting the Yukawa couplings. The key point to note here is that 
one can assign different $B$ and $L$~charges for$ H_{L,R}$ 
consistent with the above charge relation. For example, $H_L(2,1,1/2,-1/2)$, $H_R(1,2,1/2,-1/2)$ is a viable choice 
for spontaneous symmetry breaking of LB-LRSM to SM. Similarly, the spontaneous symmetry breaking of LB-LRSM to SM 
can be implemented with scalar triplets $\Delta_L(3,1,1,-1) \oplus \Delta_R(1,3,1,-1)$. In those classes of LB-LRSM theories,  
the neutral component 
of the fermion triplets, originally introduced for gauge anomaly cancellation, can be a viable dark matter candidate 
where stability is ensured automatically. Since both Higgs doublets as well as 
Higgs triplets do not couple to leptons, there is no Majorana mass term for left-handed as well as right-handed neutrinos, and thus, type-I 
plus type-II seesaw terms are absent. Then, the mechanism of the light neutrino masses can be understood via extended seesaw 
mechanism by adding singlet fermions which is beyond 
the scope of this work.
\section{Stable dark matter  Candidates}
The prime aim here is to present viable dark matter candidates within 
lepto-baryonic left-right symmetric theories. 
\subsection{Lepto-Baryons as dark  matter }
From previous discussions, it is clear that we need lepto-baryons [which transform under $SU(2)$ as triplets] 
$\Sigma_L(3,1,-3/4,-3/4)\oplus \Sigma_R(1,3,-3/4,-3/4)$, as a simple example, to make the LB-LRSM anomaly-free gauge theory. 
The key point to note here is that the neutral component of these lepto-baryons needed for anomaly cancellation can be a stable dark matter 
candidate where stability of the dark matter is ensured automatically. The relevant interaction Lagrangian for lepto-baryonic 
dark matter within LB-LRSM is given by
\begin{eqnarray}
\mathcal{L}_{\Sigma}&=&-\frac{y_{\Sigma}}{2} S_{LB} \mbox{Tr}\left(\Sigma_L \Sigma_L + \Sigma_R \Sigma_R \right) \nonumber \\
&&-e  A_\mu \overline{\Sigma^+_{L,R}} \gamma^\mu \Sigma^+_{L,R} - g_{L,R} c_W Z_{L,R_\mu} \overline{\Sigma^+_{L,R}} \gamma^\mu \Sigma^+_{L,R} 
\nonumber 
\end{eqnarray}
\begin{eqnarray}
&&-g_{L,R} \left(W^+_{L,R_\mu} \overline{\Sigma^+_{L,R}} \gamma^\mu \Sigma^0_{L,R}\mbox{+h.c.} \right) \nonumber \\
&&-\frac{3}{4}g_B Z_{B_\mu} \left(\overline{\Sigma^+_{L,R}} \gamma^\mu \gamma^5 \Sigma^+_{L,R} +
\frac{1}{4} \Sigma^0_{L,R} \gamma^\mu \gamma^5 \Sigma^0_{L,R}\right) \nonumber \\
&&-\frac{3}{4}g_\ell Z_{\ell_\mu} \left(\overline{\Sigma^+_{L,R}} \gamma^\mu \gamma^5 \Sigma^+_{L,R} +
\frac{1}{4} \Sigma^0_{L,R} \gamma^\mu \gamma^5 \Sigma^0_{L,R}\right)
\end{eqnarray}
The radiative corrections give mass splitting between neutral and charged components of lepto-baryonic fermion 
triplets. For left-handed fermion triplets, the mass splitting is $166$~MeV, whereas the mass splitting 
among right-handed triplets will depend upon the mass of the heavy gauge bosons $W_R, Z_R$ involved in the loop. 
The dark matter phenomenology for triplets will differ from Refs.\cite{Heeck:2015qra,Garcia-Cely:2015quu} 
due to additional coannihilation effects from $Z_{B,\ell}$ gauge boson-mediated diagrams. Thus, in LB-LRSM, 
the possible annihilation channels are as follows:

\begin{itemize}
\item The scalar portal annihilation channels are
\begin{eqnarray}
\Sigma_1 \Sigma_1 &\to &  H_i^* \to \bar{q}_i q_i, \bar{e}_i e_i, \bar{\nu}_i \nu_i, \nonumber \\
&& WW, ZZ, Z^\prime_{1,2} Z^\prime_{1,2}, W_R^+ W_R^-,H^{\pm \pm} H^{\mp \mp}, \nonumber 
\end{eqnarray}
Here, we denote $H_i$ and $H^{\mp \mp}$ as the heavy scalars and $Z^\prime_{1,2}$ as extra neutral gauge bosons.  
We can ignore most of these channels by assuming the small mixing between the singlet scalar $S_{LB}$ 
and other Higgs scalars. 
\item The $Z^\prime$ portal annihilation channels are 
$$\Sigma_1 \Sigma_1 \to Z_i^{'} \to \bar{q}_i q_i, \bar{e}_i e_i, \bar{\nu}_i \nu_i.$$
\item 
The important t-channel processes contributing to relic density calculation are
$$\Sigma_1 \Sigma_1 \to H_i Z_j^\prime, Z_{i}^\prime Z_j^{'}, H_i H_j, W^+ W^-, W_R^+ W_R^-, H^{\pm} W^{\mp}.$$
\end{itemize}

Following the dark matter phenomenology discussed in Refs.\cite{Heeck:2015qra,Garcia-Cely:2015quu} and the standard calculation 
for Majorana dark matter via the $Z_{B,\ell}$-portal, one can explain PLANCK-WMAP\cite{Ade:2015xua} with a few TeV dark matter masses. 
The detailed dark matter phenomenology is beyond the scope of this paper. 

\subsection{Minimal Left-Right dark  matter  possibilities}
Here the possibilities of having stable TeV-scale dark  matter  in lepto-baryonic left--right models 
is explained in a manner followed in a recent work \cite{Heeck:2015qra} 
where the stability of the dark matter is either ensured automatically because of a high $SU(2)$~dimension or due to the remnant 
discrete symmetry arising after spontaneous breaking of the $U(1)_{B-L}$ gauge group. 
The goal of this work is to only provide all stable dark matter components, and omit a detailed 
discussion, as the same has been discussed in \cite{Heeck:2015qra,Garcia-Cely:2015quu}.

At first the scalar singlet $S_{LB}(1,1,3/2,3/2)$ under $SU(2)_{L,R}$ breaks lepto-baryonic left-right 
theories $SU(2)_L \times SU(2)_R \times U(1)_B \times U(1)_L$ to conventional left-right models 
$SU(2)_L \times SU(2)_R \times U(1)_{B-L}$. Later the spontaneous breaking of LRSM to SM and 
further breaking of SM to a low-energy theory is governed by usual scalar bidoublet $\Phi(2,2,0)$ and scalar triplets $\Delta_R(1,3,2), \Delta_L(3,1,2)$. 
When the symmetry breaking $SU(2)_R \times U(1)_{B-L}$ occurs at conventional left-right breaking scale, there is a remnant discrete symmetry 

$\mathcal{Z}_2 \simeq (-1)^{B-L} \, \mbox{or\,} (-1)^{3(B-L)}$
\cite{Heeck:2015qra,Garcia-Cely:2015quu}.

Under this remnant discrete symmetry, all quarks and leptons are odd while all bosons are even. Thus, one can add a new dark matter multiplet 
which is a fermion (scalar) transforming under remnant discrete symmetry $\mathcal{Z}_2$ as even (odd). Also, the multiplets having a high 
$SU(2)$~dimension can act as a dark matter candidate if the tree level decay is forbidden at the renormalizable level. Thus, the different 
dark matter multiplets are presented below where the new dark matter particle is accidentally stable because the high $SU(2)$ 
dimensionality forbids renormalizable couplings that could lead to decay, or the stability of dark matter is ensured due to their quantum numbers 
under the remnant discrete symmetry $\mathbb{Z}_2\subset U(1)_{B-L}$ arising after spontaneous breaking of the left-right gauge group via 
the scalar triplets $\Delta_{L,R}$.

{\bf Fermionic dark  matter :}\,
\begin{align}
 \chi_L \sim (\vec{2 n +1},\vec{1},0)\,, &&
 \chi_R \sim (\vec{1},\vec{2 n +1},0)\,,
\label{eq:fermion_multiplets}
\end{align}
where $n\in \mathbb{N}$ and this results in two component dark matter with common 
Majorana mass $M$ due to LR exchange symmetry. The condition, $B-L=0$ 
for these fermionic dark matter multiplets $(\vec{2 n +1},\vec{1},0)+(\vec{1},\vec{2 n +1},0)$ 
can be contained in a higher multiplet of LB-LRSM either with the same or zero $B$ and $L$ charges. This possibility 
may give different dark matter phenomenology unlike in minimal left-right dark matter 
phenomenology \cite{Heeck:2015qra,Garcia-Cely:2015quu}. 

{\bf Scalar dark  matter :}\,
For dark matter multiplets whose stability is ensured automatically because of higher 
$SU(2)~$dimension forbidding any tree level decay of dark matter,
\begin{align}
&\chi_L (\vec{2j_1+1},\vec{2j_2+1},0 )\oplus \chi_R (\vec{2j_2+1},\vec{2j_1+1},0) \nonumber \\
& \mathrm{or} \quad \eta (\vec{2j+1},\vec{2j+1},0 )\,,
\label{eq:real_scalar_rep}
\end{align}

In addition, there could be other dark matter possibilities whose stability is guaranteed 
by the remnant discrete symmetry $\mathcal{Z}_2$ after spontaneous symmetry breaking of $SU(2)_R \times 
U(1)_{B-L}$
with $j,j_{1,2}\in \mathbb{N}$. The simplest example is scalar doublet dark matter 
\begin{align}
H_L(\vec{2},\vec{1}, -1 )\oplus H_R(\vec{1},\vec{2},-1) \,.
\end{align}

\subsection{Other dark  matter  possibilities}

{\bf Fermion Singlet Dirac dark  matter :}\,
Very recently, it was pointed out in Ref.\cite{Patra:2015vmp} that a vectorlike fermionic Dirac dark matter 
can be easily accommodated within LB-LRSM. The relevant interactions for this singlet fermion Dirac dark matter 
$\chi(1,1,n_b, n_\ell)$ with $n_B=n_\ell$ is
\begin{align}
\mathcal{L}^{\rm Dirac}_{\chi}&=-n_b g_B \overline{\chi} \gamma_\mu  \chi Z^\mu_{B}
-n_\ell g_\ell \overline{\chi} \gamma_\mu  \chi Z^\mu_{\ell}  \nonumber \\
&-\frac{1}{3} g_B \overline{q} \gamma_\mu q Z^\mu_{B}
- g_\ell \overline{\ell} \gamma_\mu \ell Z^\mu_{\ell}   \nonumber \\
&-M_\chi \overline{\chi} \chi +\cdots 
\end{align}

{\bf Fermion Singlet Majorana dark  matter :}\,
Include a Majorana fermion singlet under $SU(2)_{L,R}$ and 
equal baryon and lepton number  $\chi_L(1,1,n_{b}, n_\ell)\oplus \chi_R (1,1,n_b, n_\ell)$ 
to the minimal setup for lepto-baryonic left-right symmetric theory. The relevant interactions 
for Majorana dark  matter  is given by
\begin{align}
\mathcal{L}^{\rm Majorana}_{\chi}&- \frac{3}{4}  g_B \overline{\chi_L} \gamma_\mu \gamma^5 \chi_L Z^\mu_{B}
-\frac{3}{4} g_\ell \overline{\chi_L} \gamma_\mu \gamma^5 \chi _L Z^\mu_{\ell}  \nonumber \\
&-\frac{3}{4} g_B \overline{\chi_R} \gamma_\mu \gamma^5 \chi_R Z^\mu_{B}
-\frac{3}{4} g_\ell \overline{\chi_R} \gamma_\mu \gamma^5 \chi _R Z^\mu_{\ell}  \nonumber \\
&-\frac{1}{3} g_B \overline{q} \gamma_\mu q Z^\mu_{B}
- g_\ell \overline{\ell} \gamma_\mu \ell Z^\mu_{\ell}   \nonumber \\
&-\lambda_{\chi} S_{LB} \overline{\chi^c}_L \chi_L-\lambda'_{\chi} S_{LB} \overline{\chi^c}_R \chi_R
\end{align}
The phenomenology of Majorana fermion dark matter via $Z^\prime$ portal as well 
as Higgs portal will be studied in a separate work.

{\bf Fermion Doublet dark  matter :}\,
A few solutions for gauge anomaly cancellation within lepto-baryonic LR theories 
are presented in the Appendix. One of these solutions is lepto-baryons having doublet under $SU(2)_{L,R}$ 
with lepton and baryon number as $-3$. This immediately results in $B-L=0$, and the electric charge relation
$$Q=T_{3L}+T_{3R}+\frac{B-L}{2}$$
implies the neutral component has a nonzero hypercharge. Thus, the fermionic doublet dark matter can couple to SM $Z$ 
and thereby is ruled out from direct detection constraints.

\section{Conclusion}
The lepto-baryonic Left-Right symmetric theory is addressed by defining leptons and baryons 
as local gauge symmetries. The one-stage and two-stage spontaneous breaking of this lepto-baryonic 
left-right symmetric theory to the SM gauge theory via scalar triplets and scalar doublets are studied. 
The viability of stable dark matter candidates as triplets, doublets and singlets (Dirac as well as
Majorana fermion) in these gauge anomaly-free theories is also discussed. The detailed numerical dark matter phenomenology 
has been omitted for a separate work. Towards the end the Lagrangian interaction for both triplet and doublet versions 
of lepto-baryonic left-right symmetric theories has been provided.

\section{Acknowledgement} 
The author thanks Werner Rodejohann for the visiting position at the Max-Planck Institute for Kernphysik, Heidelberg, Germany 
where this idea was conceived. The work of SP is partially supported by the Department of Science and Technology, Govt.\ of India 
under the financial grant SB/S2/HEP-011/2013.

\section{Appendix: Anomalies in Left-Right theories with $SU(2)_L \times SU(2)_R \times U(1)_{B} \times U(1)_L$}
We discuss below the induced gauge anomalies in the left-right symmetric model with the gauge group 
$SU(2)_L \times SU(2)_R \times U(1)_{B} \times U(1)_L$ omitting the $SU(3)_C$ structure for simplicity 
where baryon and lepton numbers are promoted to local gauge symmetries. For completeness, we also 
present a complete Lagrangian for a simplified version of this class of left-right symmetric theories. 
\subsection{Few gauge anomaly free solutions for lepto-baryonic Left-Right Models}
The lepto-baryonic Left-Right symmetric models with gauge group
$\mathcal{G}^{BL}_{L,R}\equiv SU(2)_L \times SU(2)_R \times U(1)_{B} \times U(1)_{L}$--
skipping the $SU(3)_C$ structure for simplicity-- induces non-zero gauge anomalies in the theory, 
\begin{align}
&\mathcal{A}\left[ SU(2)^2_L\times U(1)_B\right]=3/2\, , \nonumber \\
&\mathcal{A}\left[ SU(2)^2_R\times U(1)_B\right]=-3/2\, , \nonumber \\
&\mathcal{A}\left[ SU(2)^2_L\times U(1)_L\right]=3/2\, , \nonumber \\
&\mathcal{A}\left[ SU(2)^2_R\times U(1)_L\right]=-3/2\, , \nonumber 
\end{align}
Although solutions to these gauge anomalies have been presented in Refs.\cite{He:1989mi, Patra:2015vmp}, it will be good 
to note them down here so that dark  matter studies can be applicable to other scenarios also, if applicable, as 
\begin{itemize}
\item The simplest way to cancel a gauge anomaly is to include the fermion triplets $\Sigma_L\oplus \Sigma_R$ 
as follows:
\begin{eqnarray}
\Sigma_{L}=\begin{pmatrix}
  \Sigma^0_{L,R}  & \sqrt{2} \Sigma^+_{L,R}  \\
  \sqrt{2} \Sigma^-_{L,R} & -\Sigma^0_{L,R}
 \end{pmatrix}\, , \quad  \Sigma_{R}=\begin{pmatrix}
  \Sigma^0_{R}  & \sqrt{2} \Sigma^+_{R}  \\
  \sqrt{2} \Sigma^-_{R} & -\Sigma^0_{R}
 \end{pmatrix}\, , 
\end{eqnarray}

\item
Lepto-baryons doublet under $SU(2)_{L,R}$ and color singlet carrying nonzero baryon and lepton numbers,
\begin{eqnarray}
\mbox{$SU(3)_C$ singlets}
\bigg\{\begin{array}{cc}
        \Psi^1_L \sim (2,1,-n,-n) & \\
        \Psi^1_R \sim (1,2,-n,-n)\,,& \quad 
       \end{array}
\end{eqnarray}

\item 
Lepto-Baryons which are nonsinglets under both $SU(2)_{L,R}$ and  triplets under $SU(3)_C$ with nonzero lepton and baryon numbers, 
\begin{eqnarray}
\mbox{$SU(3)_C$ triplets}
\bigg\{\begin{array}{cc}
         \Psi^3_L \sim (2,1,-n/3,-n/3) & \\
        \Psi^3_R \sim (1,2,-n/3,-n/3)\,,& 
       \end{array}
\end{eqnarray}
\item  Lepto-baryons with $N-$dimensional representation of 
$SU(3)_C$ and doublets under $SU(2)_{L,R}$ with nonzero $B$ and $L$, 
\begin{eqnarray}
	\mbox{$SU(3)_C$ N-tuplets}
\bigg\{\begin{array}{c}
        \Psi^N_L \sim (2,1,-n/N,-n/N) \\
        \Psi^N_R \sim (1,2,-n/N,-n/N)
       \end{array}
\end{eqnarray}
with $n$ being the number of fermion family generation which is 3,
\end{itemize}

\subsection{Lagrangian for lepto-baryonic Left-Right Theories}
The Lagrangian for the present lepto-baryonic left-right theories where the spontaneous 
symmetry breaking is achieved with scalar sector comprising of $H_R, H_L, \Phi, S_{LB}$ 
is given below:
\begin{eqnarray}
\mathcal{L}^{\rm LB}_{\rm LR}&=& \mathcal{L}^{\rm }_{\rm scalar} + \mathcal{L}^{\rm gauge}_{\rm Kin.} 
      + \mathcal{L}^{\rm fermion}_{\rm Kin.} + \mathcal{L}_{\rm Yuk}
\end{eqnarray}
where the individual parts can be written as
\begin{eqnarray}
\mathcal{L}^{\rm }_{\rm scalar} &=&
      \mbox{Tr}\big[\left(\mathcal{D}_\mu \Phi\right)^\dagger \left(\mathcal{D}^\mu \Phi\right) \big] 
      +\left(\mathcal{D}_\mu S_{BL}\right)^\dagger \left(\mathcal{D}^\mu S_{BL}\right)  \nonumber \\
      &+& \left(\mathcal{D}_\mu H_L\right)^\dagger \left(\mathcal{D}^\mu H_L\right)
      + \left(\mathcal{D}_\mu H_R\right)^\dagger \left(\mathcal{D}^\mu H_R\right) \nonumber \\
      && - \mathcal{V}(\Phi, H_L, H_RS_{BL}) 
\end{eqnarray} 
Defining $\Phi \equiv \Phi_1$ and $\Phi_2=\tau_2 \Phi^* \tau_2$, the scalar potential can be written as follows
\begin{eqnarray}     
&&\hspace*{-0.5cm} \mathcal{V}(\Phi, H_L, H_R, S_{LB})=
-\sum_{i,j=1,2} \frac{\mu_{\phi ij}^{2}}{2}~\mbox{Tr}(\Phi_{i}^{\dagger}\Phi_{j}) \nonumber \\
   && +\sum_{i,j,k,l=1,2}\frac{\lambda_{\phi ijkl}}{4}~\mbox{Tr}(\Phi_{i}^{\dagger}\Phi_{j})~\mbox{Tr}(\Phi_{k}^{\dagger}\Phi_{l}) 
    \nonumber \\
   &&+\sum_{i,j,k,l=1,2}\frac{\Lambda_{\phi ijkl}}{4}~\mbox{Tr}(\Phi_{i}^{\dagger}\Phi_{j}\Phi_{k}^{\dagger}\Phi_{l}) \nonumber \\
&&     -\mu^2_H\, \left(H^\dagger_L H_L +H^\dagger_R H_R \right) \nonumber \\
   &&+\lambda_1\, \left[\left(H^\dagger_L H_L\right)^2+ \left(H^\dagger_R H_R \right)^2 \right]
     +\lambda_2\, \left(H^\dagger_L H_L\right) \left(H^\dagger_R H_R \right)
   \nonumber \\
   &&+\sum_{i,j} \beta_{ij}~\left(H^\dagger_L H_L +H^\dagger_R H_R \right)\mbox{Tr}(\Phi_{i}^{\dagger}\Phi_{j})  \nonumber \\
&&    +\sum_{i,j} \varrho_{ij}\left[H^\dagger_L \Phi_i \Phi^\dagger_j H_L+H^\dagger_R \Phi^\dagger_i \Phi_i H_R\right]
    \nonumber \\
   &&-\mu^2_S S^\dagger_{LB} S_{LB} + \lambda_S \left(S^\dagger_{LB} S_{LB}\right)^2  \nonumber \\
  &&   +\lambda_{SH}\,\left(S^\dagger_{LB} S_{LB}\right) \left(H^\dagger_L H_L +H^\dagger_R H_R \right)
       \nonumber \\
   &&+\sum_{i,j}\lambda_{S\phi ij}\,\,\left(S^\dagger_{LB} S_{LB}\right) \mbox{Tr}(\Phi_{i}^{\dagger}\Phi_{j})\, .
\end{eqnarray}
The kinetic terms for the gauge bosons are given by
\begin{eqnarray} 
\mathcal{L}^{\rm gauge}_{\rm Kin.}&=&
     -\frac{1}{4}W_{\mu\nu L}.W^{\mu\nu L}-\frac{1}{4}W_{\mu\nu R}.W^{\mu\nu R}   \nonumber \\
   &&     -\frac{1}{4}B^B_{\mu\nu}B^{\mu\nu,B} -\frac{1}{4}B^L_{\mu\nu}B^{\mu\nu,L}
\end{eqnarray}
while for fermions,
\begin{eqnarray} 
\mathcal{L}^{\rm fermion}_{\rm Kin.}&=&
       i  \overline{q_{L}}\gamma^{\mu} \mathcal{D}_\mu q_{L}
      +i  \overline{q_{R}}\gamma^{\mu} \mathcal{D}_\mu q_{R}  \nonumber \\
    &&  +i  \overline{\ell_{L}}\gamma^{\mu} \mathcal{D}_\mu \ell_{L}
      +i  \overline{\ell_{R}}\gamma^{\mu} \mathcal{D}_\mu \ell_{R}
      \nonumber \\
    &&+i\overline{\Sigma_{L}}\gamma^{\mu} \mathcal{D}_\mu \Sigma_{L}
      +i  \overline{\Sigma_{R}}\gamma^{\mu} \mathcal{D}_\mu \Sigma_{R}
\end{eqnarray}
However we can define the respective covariant derivatives, in general, as
\begin{eqnarray} 
\mathcal{D}^f_\mu=\partial_{\mu}-i\,g_L \tau^a W^a_{\mu L} -i\,g_R \tau^a W^a_{\mu R}
                  -i\,g_B\frac{B}{2} B^B_{\mu}-i\,g_L\frac{L}{2} B^L_{\mu} \nonumber
\end{eqnarray}
The Yukawa interaction Lagrangian can be read as
\begin{eqnarray}
\mathcal{L}_{\rm Yuk}&=&
     Y_q\, \overline{Q_{L}} \Phi Q_{R} + \widetilde{Y_q} \overline{Q_{L}}  \widetilde{\Phi}\, Q_{R} \nonumber \\
   &&     +Y_\ell\, \overline{\ell_{L}} \Phi \ell_{R} + \widetilde{Y_\ell} \overline{\ell_{L}} \widetilde{\Phi}\, \ell_{R}\nonumber \\
     &&+\lambda_\Sigma \left(\Sigma^T_L C \Sigma_L +\Sigma^T_R C \Sigma_R \right)\,S_{LB}+ \mbox{h.c.}\,
\end{eqnarray}

\bibliographystyle{utcaps_mod}
\bibliography{onubb_LR}
\end{document}